\documentclass[12pt]{article}
\usepackage{epsfig}
\usepackage{amssymb}
\usepackage{amsmath}
\usepackage{amsfonts}

\newcommand{\tr}{\mathrm{Tr}}

\newcommand{\tI}{\mathrm{I}}
\newcommand{\tII}{\mathrm{II}}

\def\RR{\mathbb{R}}

\newcommand{\cC}{\mathcal{C}}
\newcommand{\cD}{\mathcal{D}}

\newcommand{\cI}{\mathcal{I}}
\newcommand{\cJ}{\mathcal{J}}

\newcommand{\cP}{\mathcal{P}}

\newcommand{\cT}{\mathcal{T}}

\newcommand{\be}{\begin{equation}}
\newcommand{\ee}{\end{equation}}
\newcommand{\bea}{\begin{eqnarray}}
\newcommand{\eea}{\end{eqnarray}}
\newcommand{\nn}{\nonumber}
\newcommand{\kt}{\rangle}
\newcommand{\br}{\langle}

\newcommand{\ed}{\end{document}}

\newcommand{\bi}{\begin{itemize}}
\newcommand{\ei}{\end{itemize}}

\newcommand{\bce}{\begin{center}}
\newcommand{\ece}{\end{center}}

\begin{document}

\title{Pseudo-Entanglement Evaluated in Noninertial Frames}

\author{Hossein Mehri-Dehnavi\thanks{E-mail address:
mehri@alice.math.kindai.ac.jp}, Behrouz Mirza\thanks{E-mail address:
b.mirza@cc.iut.ac.ir}, Hosein Mohammadzadeh\thanks{E-mail address:
h.mohammadzadeh@ph.iut.ac.ir},\\ and Robabeh Rahimi\thanks{E-mail
address: rrahimid@uwaterloo.ca}
\\
\\
$^*$~Department of Physics, Institute for Advanced Studies in Basic
Sciences,\\ Zanjan 45195-1159, Iran,\\
$^*$~Research Center for Quantum Computing, Kinki University,
3-4-1 Kowakae,\\ Higashi-Osaka, Osaka 577-8502, Japan\\
$^{\dagger},^{\ddagger}$~Department of Physics, Isfahan University
of Technology,\\ Isfahan 84156-83111, Iran
\\
$^{\S}$~Department of Chemistry, Graduate School of Science, Osaka
City University, \\Osaka 558-8585, Japan\\
$^{\S}$~Institute for
Quantum Computing and Department of Physics and Astronomy,\\
University of Waterloo, Waterloo, ON, N2L 3G1, Canada.}
\date{ }
\maketitle

\begin{abstract}
We study quantum discord, in addition to entanglement, of bipartite
pseudo-entanglement in noninertial frames. It is shown that the
entanglement degrades from its maximum value in a stationary frame
to a minimum value in an infinite accelerating frame. There is a
critical region found in which, for particular cases, entanglement
of states vanishes for certain accelerations. The quantum discord of
pseudo-entanglement decreases by increasing the acceleration. Also,
for a physically inaccessible region, entanglement and nonclassical
correlation are evaluated and shown to match the corresponding
values of the physically accessible region for an infinite
acceleration.

 \end{abstract}

\vspace{5mm}


\maketitle
\section{Introduction}
Implementations of quantum information protocols and achievements of
any advantage over classical computers have formed the underlying
causes for the high popularity of researches on  quantum information
processing. This is generally true for studies not only in an
inertial frame but also for those within  noninertial frames when
the relative motion and/or the acceleration of the communicating
partners are large \cite{ref01}. Quantum entanglement, mentioned as
the essence of quantum physics \cite{ref02} and believed to be a
possible quantum resource for a quantum processor in surpassing the
presently available classical computers, has been widely studied
theoretically \cite{ref03} and experimentally, for a variety of
physical systems as well as for inertial/noninertial frames
\cite{ref01,ref04,ref05,ref06,ref07,ref08,ref09,ref10,ref11,ref12,ref13,ref14,ref15,ref16}.

When entanglement is considered in noninertial frames, a
relativistic frame should be employed.  In principle, an observer
with uniform acceleration cannot obtain information about the whole
spacetime from his perspective, which leads to a communication
horizon to appear. This results in a loss of information and a
corresponding degradation of entanglement. It has been shown that
the Unruh effect \cite{ref17} degrades the entanglement between
partners \cite{ref05,ref06,ref07}. Therefore, the implementation of
certain quantum information processing tasks between accelerated
partners requires a quantitative understanding of such degradation
in noninertial frames. As a natural question, one may ask how the
involved parties in a state, particularly in an entangled state,
behave in accordance to each other from the viewpoints of the
accelerating and resting observers.

The key role of entanglement for quantum information processing, in
general, has not been yet proved. The power of a quantum processor
has been also mentioned to be due to any existence of correlation
other than the purely classical one. Such correlations include, but
not restricted, the entanglement of the states. Meanwhile, attempts
have been made to revise the distinct role of entanglement since the
original work by Ollivier and Zurek \cite{ref18}, stating that a
separable state (hence nonentangled by definition) may present
quantum correlation \cite{ref19,ref20}.

Nonclassically correlated states other than entanglement have shown
to present computational powers beyond classical schemes
\cite{ref21}. In addition, almost all the implementations of quantum
information protocols by conventional nuclear magnetic resonance
(NMR) \cite{ref22}, electron nuclear double resonance (ENDOR)
\cite{ref23}, or other similar bulk ensemble quantum computers at
room temperature have been under experimental conditions of what
just could provide a pseudo-pure state as follows
    \bea
    \rho_{\rm ps}=\frac{1-p}{4}I+p|\psi\kt\br\psi|,
    \label{pseudo-gen}
    \eea
where, $p$ characterizes the fraction of state $|\psi\kt$ which has
been originally the desired initial simple and fiducial  state for
quantum computation, $|00\kt$ \cite{ref24}. In order to produce a
particular entangled state, the corresponding entangling operation
$E$ is applied to the pseudo-pure state and changes it to
    \bea
    \rho=\frac{1-p}{4}I+p|\Phi^+\kt\br\Phi^+|,
    \label{pseu-ent-gen}
    \eea
where, $|\Phi^+\kt=E|\psi\kt=\frac{1}{\sqrt{2}}(|00\kt+|11\kt)$.
This state is entangled if and only if $p>1/3$. Otherwise, for a
case of $p> 0$, it is separable but yet involves nonclassical
correlations. In an experiment involving two spins representing to
the two parties $A$ and $B$, $E$ can be a combination of Hadamard
gate on $A$ followed by a CNOT gate on both sides.

Recent trends point toward extending quantum information processing
to noninertial frames. Also, once a bulk ensemble system is supposed
as a testbed for implementations in either inertial or noninertial
frames, the existence and, accordingly, the behavior of the
nonclassical correlations are even more important than the status of
entanglement. In view of this demand, in this work, we study the
pseudo-entangled state (\ref{pseu-ent-gen}) for noninertial frames.
This can be complementary to previous  studies for an inertial
frame. However, for the sake of selfcompleteness, results of an
overview on the existence of entanglement are given in an
appropriate place.

This paper is organized as follows. Section II is to give a short
survey of entanglement and nonclassical correlation; for each class
a measure is described which is extensively employed in this work.
Section III is a short review on the Dirac fields in noninertial
frames. Section IV provides the main results of the study. Finally,
summary and discussions will conclude the paper.

\section{Classical/Nonclassical  Correlations}
Classical/nonclassical correlations for subsystems of a quantum
system have been indispensable subjects in quantum information
theory. For historical reasons, a nonclassically correlated state
may be explained by reviewing the concept of entanglement.

An entangled state, according to separability paradigm
\cite{ref25,ref26}, cannot be prepared by local operations and
classical communications (LOCC) \cite{ref27}. For a density matrix
$\rho$ of a composite bipartite system $AB$, a separable state can
be written as follows
    \be
    \rho_{\rm sep}=\sum_iw_i \rho_A^i\otimes \sigma_B^i,
    \label{rho-sep}
    \ee
where, $w_i$'s are positive weights, and $\rho_A^i$'s and
$\sigma_B^i$'s are local states belonging to $A$ and $B$,
respectively. An entangled state $\rho_{\rm ent}$ is an inseparable
state that is not of the above form.

There are other paradigms that are mostly  based on a
post-preparation stage \cite{ref18,ref28,ref29,ref30}  rather than a
preparation one. According to Oppenheim-Horodecki paradigm
\cite{ref31}, a properly classically correlated state of a bipartite
system $AB$, $\rho_{\rm pcc}$, is defined as a state represented by
    \be
    \rho_{\rm pcc}=\sum_{i=1}^{d^A}\sum_{j=1}^{d^B}e_{ij}
    |v_A^i\kt\br v_A^i|\otimes|v^j_B\kt\br v^j_B|,
    \label{properly-CC}
    \ee
where, $d^A$ and $d^B$ are the dimensions of the Hilbert spaces of
$A$ and $B$, respectively. $e_{ij}$ is the eigenvalue of $\rho_{\rm
pcc}$ corresponding to an eigenvector $|v^i_A\kt\otimes|v^j_B\kt$.
Based on this definition, a nonclassically correlated state,
$\rho_{\rm ncc}$, is a state that cannot be described as the above
form (\ref{properly-CC}).

Several  approaches are available for evaluating the status of
entanglement in a system \cite{ref32,ref33}. Logarithmic Negativity
of the state $\rho$, $N(\rho)$, is a nonconvex entanglement monotone
which gives one of the most powerful means for quantifying
entanglement. For a bipartite system $\rho$, $N(\rho)$, is defined
as
    \bea
    N(\rho):
    =\log_2\sum_i|\lambda_i(\rho^{\rm pt})|,
    \label{logneg}
    \eea
where, $\lambda_i(\rho^{\rm pt})$ denotes the eigenvalues of the
partial transpose, $\rho^{\rm pt}$, of $\rho$. According to the
properties of the partial transpose, the logarithmic negativity is
symmetric with respect to $A$ and $B$. This behavior is not observed
for quantum discord explained below.

As for nonclassical correlations, among the exiting measures
\cite{ref18,ref29,ref31,ref34,ref35,ref36,ref37,ref38,ref39,ref40},
we evaluate  quantum discord \cite{ref18}, which has been studied
for quantum computing \cite{ref21,ref37,ref41,ref42} and
broadcasting of quantum states \cite{ref43,ref44}. The dynamics of
quantum discord is also studied \cite{ref45}.

Before we embark on reformulating  quantum discord, it is important
to note that measures of entanglement and classical/quantum
correlations can be conceptually different from each other. For
example, since quantum discord is not generally identical with an
entanglement measure, any direct comparison of the two notions can
be meaningless. For a comparison to be valid and to produce
reasonable data, one needs to employ a unified approach
\cite{ref41}.

Quantum discord for a bipartite quantum system is the discrepancy
between the quantum mutual information and the locally accessible
mutual information. These two concepts for measuring the mutual
information context are classically identical. The quantum mutual
information is defined as
    \bea
    \cI(A:B)=S(\rho_{A})+S(\rho_{B})-S(\rho),
    \label{mutual}
    \eea
where, $\rho_A$ and $\rho_B$ are the  density operators of  $A$ and
$B$, respectively. Here, $S(\rho)=-\tr(\rho\log_2\rho)$ is the von
Neumann entropy.

The locally accessible mutual information, $\cJ$, is represented by
    \bea
    \cJ_{\{\Pi_k\}}(A:B)=S(\rho_{A})-S_{\{\Pi_k\}}(A|B).
    \label{J}
   \eea
Here, $\{\Pi_k\}$'s are von Neumann operators acting on subsystem
$B$ and corresponding to the outcome $k$. $S_{\{\Pi_k\}}(A|B)$ is
the quantum conditional entropy, that is
    \bea
    S_{\{\Pi_k\}}(A|B)=\sum_k p_k S(\rho_{A|k}),
    \label{S(A:B)}
    \eea
where, $\rho_{A|k}=\tr_B(\Pi_k \rho \Pi_k)/p_k$, with $p_k=\tr(\Pi_k
\rho \Pi_k)$. The classical correlation is given by
\cite{ref18,ref19,ref20,ref36,ref46}
     \bea
    \cC(A:B)=\max_{\{\Pi_k\}} \left[ \cJ_{\{\Pi_k\}}(A:B)\right].
    \label{CC}
    \eea
By substituting the appropriate values, the quantum discord is
calculated as
     \bea
    \cD(A:B)=\cI(A:B)-\cC(A:B).
    \label{discord}
    \eea

One can also select a different set of von Neumann operators,
$\{\Pi_k'\}$, acting on the subsystem $A$ and corresponding to the
outcome $k$, that can be written as  $S_{\{\Pi_k'\}}(B|A)=\sum_k
p_k' S(\rho'_{k|B})$; where, $\rho'_{k|B}=\tr_A(\Pi_k' \rho
\Pi_k')/p_k'$, with $p_k'=\tr(\Pi_k' \rho \Pi_k')$. Finding the
minimum value of $S_{\{\Pi_k'\}}(B|A)$, over all von Neumann
operators, we can evaluate the classical correlation as follows:
    \bea
    \cC(B:A)=S(\rho_B) -\min_{\{\Pi_k'\}}S_{\{\Pi_k'\}}(B|A).
    \label{CC2}
    \eea
Thus, quantum discord $\cD(B:A)$, is  obtained as,
     \bea
    \cD(B:A)=\cI(B:A)-\cC(B:A).
    \label{discord2}
    \eea
It has been mentioned by  Zurek \cite{ref47} that quantum discord is
not generally symmetric, {\it i.e.}, we have
    \bea
    \cD(A:B)\neq\cD(B:A).
    \label{d(AB-BA)}
    \eea
In following we will observe this property.
\section{Dirac Fields in Noninertial Frames}
Let us suppose a bipartite system. The resting part is named as
Alice, A, and the accelerating part as Rob, R. Alsing {\it et. al.} \cite{ref06} have studied the degradation of entanglement by Unruh effect for the bipartite system. If we consider Rob to
be uniformly accelerated in the $(t,z)$ plane, then the appropriate
Rindler coordinates $(\tau,\zeta)$ introduce two different Rindler
regions that are causally disconnected from each other:
 \bea
 \label{MF}&&at=e^{a\zeta}\sinh(a\tau),~~~az=e^{a\zeta}\cosh(a\tau),\nn\\
 &&at=-e^{a\zeta}\sinh(a\tau),~~~az=-e^{a\zeta}\cosh(a\tau).
 \eea
The above set of coordinates is related to regions $\mathrm{I}$ and
$\mathrm{II}$, respectively,
 and $a$ denotes Rob's proper acceleration.
One can quantize a free spinor field, in  $3+1$ dimensions, which
satisfies the Dirac equation by using the complete orthonormal set
of fermion, $\psi^{+}_{k}$, and antifermion, $\psi^{-}_{k}$, modes,
 \bea
 \psi=\int (a_{k}\psi^{+}_{k}+b^{\dag}_{k}\psi^{-}_{k})d k,
 \eea
where, $a^{\dag}_{k}$ ($b^{\dag}_{k}$) and $a_{k}$ ($b_{k}$) are the
creation and annihilation operators for fermions (antifermions) of
the momentum $k$, respectively. They satisfy the anticommutation
relation
 \bea
 \{a_{i},a^{\dag}_{j}\}=\{b_{i},b^{\dag}_{j}\}=\delta_{ij}.\label{AC}
 \eea
The  quantum field theory for the Rindler observer is constructed by
expanding the spinor field in terms of a complete set of fermion and
antifermion modes in regions $\mathrm{I}$ and $\mathrm{II}$, as
already introduced in Eq. (\ref{MF}),
 \bea
 \psi=\int\sum_{\sigma}(c^{\sigma}_{k}\psi^{\sigma+}_{k}+
 d^{\sigma\dag}_{k}\psi^{\sigma-}_{k})dk,~~~~~\sigma\in\{\mathrm{I},\mathrm{II}\},
 \eea
where, $c^{\sigma\dag}_{k}$ ($d^{\sigma\dag}_{k}$) and
$c^{\sigma}_{k}$ ($d^{\sigma}_{k}$) are the creation and
annihilation operators for fermion (antifermions), respectively,
acting on region $\mathrm{I}$ ($\mathrm{II}$) for
$\sigma=\mathrm{I}$ ($\mathrm{II}$) and satisfying  an
anticommutation relation similar to Eq. (\ref{AC}). One can find a
relation between creation and annihilation operators of Minkowski
and Rindler spacetime using Bogoliubov transformation
\cite{ref48,ref49,ref50}
 \bea
 a_{k}=\cos r c^{\mathrm{I}}_{k}-\sin r
 d^{\mathrm{II}\dag}_{-k},~~~~~~
 b_{k}=\cos r d^{\mathrm{I}}_{k}+\sin r c^{\mathrm{II}\dag}_{-k},\label{BT}
 \eea
where, $\cos r=1/\sqrt{1+e^{-2\pi \omega c/a}}$ with
$\omega=\mbox{$\sqrt{|\vec{k}|^2+m^2}$}$. It is easy to see from Eq.
(\ref{BT}) and its adjoint that Bogoliubov transformation mixes a
fermion in region $\mathrm{I}$ and antifermions in region
$\mathrm{II}$. Therefore, we postulate that the Minkowski particle
vacuum state for mode $k$ in terms of Rindler Fock states is given
by
 \bea
 |0_{k}\kt_{M}=\sum_{n=0}^{1}A_{n}|n_{k}\kt^{+}_{\tI}|n_{-k}\kt^{-}_{\tII}.\label{Vd}
 \eea
As a comment on notation, the Rindler region $\mathrm{I}$ or
$\mathrm{II}$ Fock states carry a subscript $\tI$ and $\tII$,
respectively, on the kets, while the Minkowski Fock states are
indicated by the subscript $M$ on the kets. In the following, we use
the single mode approximation and we will drop all labels $(k,-k)$
on states and density matrices indicating the specific mode. By
applying the creation and annihilation operators to  Eq. (\ref{Vd})
and using the normalization condition, one can obtain \cite{ref06}
 \be
    |0\kt_M=\cos r |0\kt_\tI|0\kt_{\tII}+\sin r |1\kt_\tI|1\kt_{\tII},
    \label{0-m-to-r}
 \ee
and in a same manner
 \be
    |1\kt_M=|1\kt_\tI|0\kt_{\tII}.
    \label{1-m-to-r}
    \ee
Also, we can show that when Rob accelerates uniformly through the
Minkowski vacuum, his detector registers a number of particles given
by
 \bea
 ^{+}\br 0|c^{\tI\dag}c^{\tI}|0\kt^{+}=\frac{1}{e^{2\pi\omega/a}+1},\label{Unruh}
 \eea
where, $\omega$ is related to the specified $k$ mode. Equation
(\ref{Unruh}) is known as the Unruh effect \cite{ref17}, which
demonstrates that Rob in region $\mathrm{I}$ detects a thermal
Fermi-Dirac distribution of particles as he passes through the
Minkowski vacuum. It is notable that more recently the Unruh effect
beyond the single mode approximation has been considered
\cite{ref51}.
\section{Discord and Entanglement of Pseudo-Entangled States in Noninertial Frames}
We would like to study the state (\ref{pseu-ent-gen}) in noninertial
frames. Using Eqs. (\ref{0-m-to-r}) and (\ref{1-m-to-r}), we find
the relating density matrix as follows
     \bea
      \rho_{A,\tI,\tII}=\frac{1}{4}\left(
                                \begin{array}{cccccccc}
                                  (1+p)\cos^2r & 0 & 0 & \frac{1+p}{2}\sin 2r & 0 & 0 & 2p \cos r & 0 \\
                                  0 & 0 & 0 & 0 &                               0 & 0 & 0 & 0 \\
                                  0 & 0 & 1-p & 0 &                               0 & 0 & 0 & 0 \\
                                  \frac{1+p}{2}\sin 2r& 0 & 0 & (1+p)\sin^2r &  0 & 0 & 2p\sin r & 0 \\
                                  0 & 0 & 0 & 0 & (1-p)\cos^2r & 0 & 0 & \frac{1-p}{2}\sin 2r \\
                                  0 & 0 & 0 & 0 & 0 & 0 & 0 & 0 \\
                                  2p \cos r & 0 & 0 & 2p\sin r & 0 & 0 & 1+p & 0 \\
                                  0 & 0 & 0 & 0 & \frac{1-p}{2}\sin 2r & 0 & 0 & (1-p)\sin^2r \\
                                \end{array}
                              \right),
    \label{rho-AIII}
    \eea
where, we have used the basis $|000\kt$, $|001\kt$, $|010\kt$,
$|011\kt$, $|100\kt$, $|101\kt$, $|110\kt$, and $|111\kt$, with
$|lmn\kt:=|l\kt_A |m\kt_{\tI}|n\kt_{\tII}$, in writing the above
matrix.

A bipartite density matrix is obtained by tracing out the modes of
one of the regions $A$, $\tI$, or $\tII$, in which logarithmic
negativity and quantum discord are applied for evaluating the status
of entanglement and nonclassical correlation, respectively. It turns
out that by tracing out each element of $A$, $\tI$, or $\tII$, the
resulting matrix form of the state is a real symmetric $X$-shaped
density matrix. Therefore, we review quantum discord for the
set of $X$-states in Appendix A.
\subsection{Bipartition Alice-Rob}
For finding the relating density matrix between Alice and Rob, the
antiRob's modes of $\rho_{A,\tI,\tII}$ are traced out, {\it i.e.},
$\rho_{A,\tI}=\tr_{\tII} (\rho_{A,\tI,\tII})$. Using the basis
$|0\kt_A|0\kt_{\tI}$, $|0\kt_A|1\kt_{\tI}$, $|1\kt_A|0\kt_{\tI}$,
and $|1\kt_A|1\kt_{\tI}$, it can be written as
     \bea
      \rho_{A,\tI}=\frac{1}{4}\left(
                          \begin{array}{cccc}
                            (1+p)\cos^2 r & 0 &  0& 2 p\cos r  \\
                            0 & 1+\sin^2 r-p\cos^2 r & 0 & 0 \\
                            0 & 0 & (1-p)\cos^2r & 0 \\
                            2 p\cos r& 0 & 0 & 1+\sin^2 r+p\cos^2 r \\
                          \end{array}
                        \right).
    \label{rho-AI}
    \eea
The eigenvalues of $\rho_{A,\tI}$ are given by
    \bea
    \lambda_{1,2}(\rho_{A,\tI})&=&\frac{1}{4}\left\{1+p\cos^2r\pm\sqrt{4p^2\cos^2r+\sin^4r}\right\},\nn\\
    \lambda_{3,4}(\rho_{A,\tI})&=&\frac{1}{4}\left\{1-p\cos^2r\pm\sin^2r\right\}.
    \eea
Thus, the von Neumann entropy of $\rho_{A,\tI}$ is calculated  as
$S(\rho_{A,\tI})=-\sum_{j=1}^4 \lambda_j\log_2\lambda_j$. Similarly,
the entropy of the reduced density matrices $\rho_A$ and
$\rho_{\tI}$ are obtained as $S(\rho_A)=1$ and
$S(\rho_{\tI})=-\frac{\cos^2r}{2}\log_2\frac{\cos^2r}{2}
-\frac{1+\sin^2r}{2}\log_2\frac{1+\sin^2r}{2}$, respectively.

\begin{figure}[t]
    \begin{center}
   \includegraphics[scale=.17,clip]{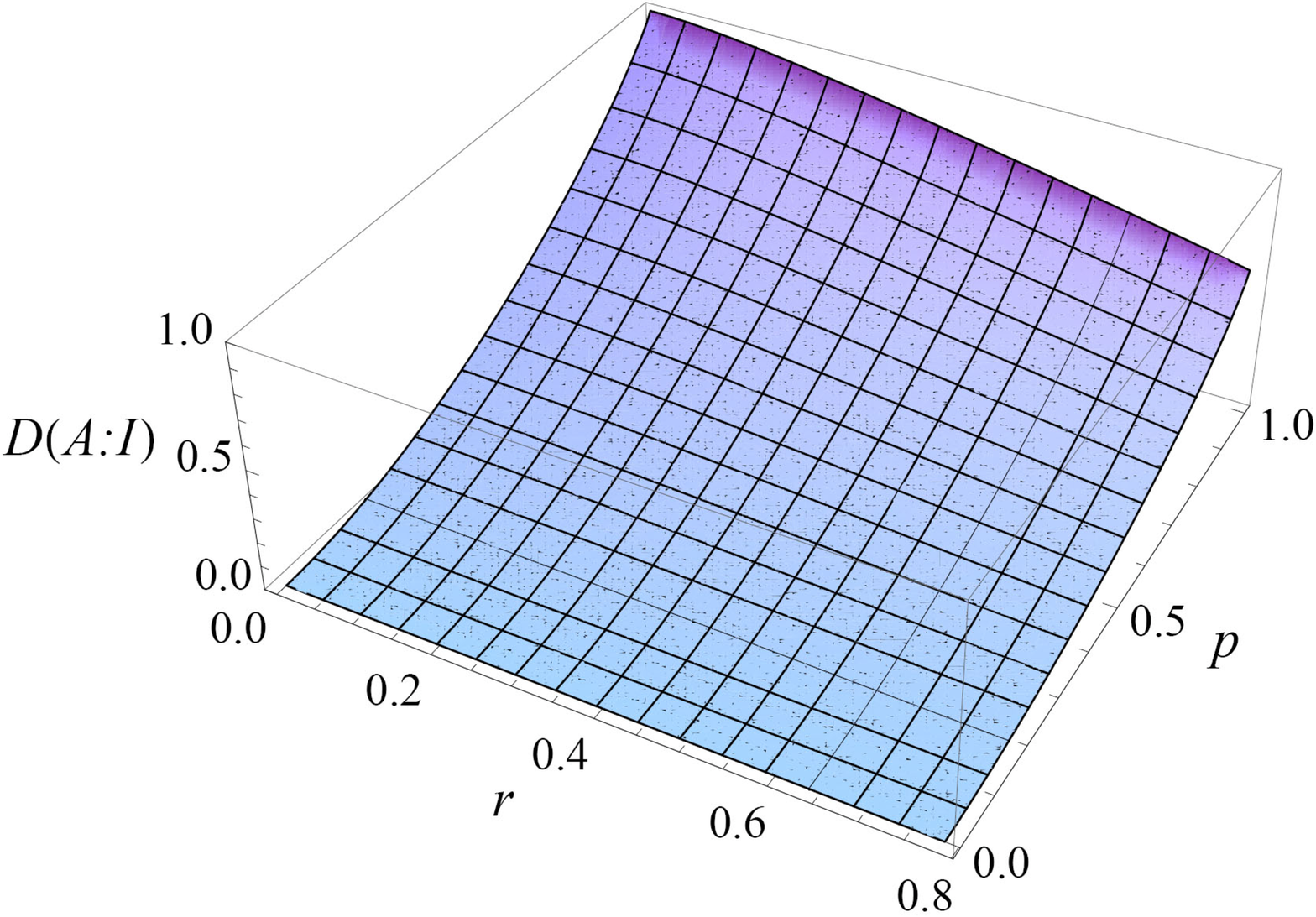}
    \parbox{15cm}{\caption{  Quantum discord, $\cD(A:\tI)$,
as function of $r$ and $p$.
    \label{fig1}}}\end{center}
    \end{figure}

Therefore, quantum discord is given by
    \bea
    \cD(A:\tI)=\min_{\{\Pi_k\}}[S_{\{\Pi_k\}}(A|\tI)]-S(\rho_{A,\tI})-
    \mbox{$\frac{\cos^2r}{2}$}\log_2\mbox{$\frac{\cos^2r}{2}$}-
    \mbox{$\frac{1+\sin^2r}{2}$}\log_2\mbox{$\frac{1+\sin^2r}{2}$}.
    \label{disAI}
    \eea
It has been  discussed in the Appendix that it suffices to simply
check four end points. By plotting $S_{\{\Pi_k\}}(A|\tI)$ for these
cases, the case $\kappa=\ell=1/2$ with $\beta=1/4$ is corresponding
to this minimum. Replacing this value in Eq. (\ref{disAI}) quantum
discord is plotted in Fig. \ref{fig1}. It is clear that quantum
discord is a decreasing function of acceleration. The important
point is that, for all inertial nonclassically correlated cases
($p>0$), it is impossible to transfer a state to a properly
classically correlated state.

\begin{figure}[t]
    \begin{center}
   \includegraphics[scale=.17,clip]{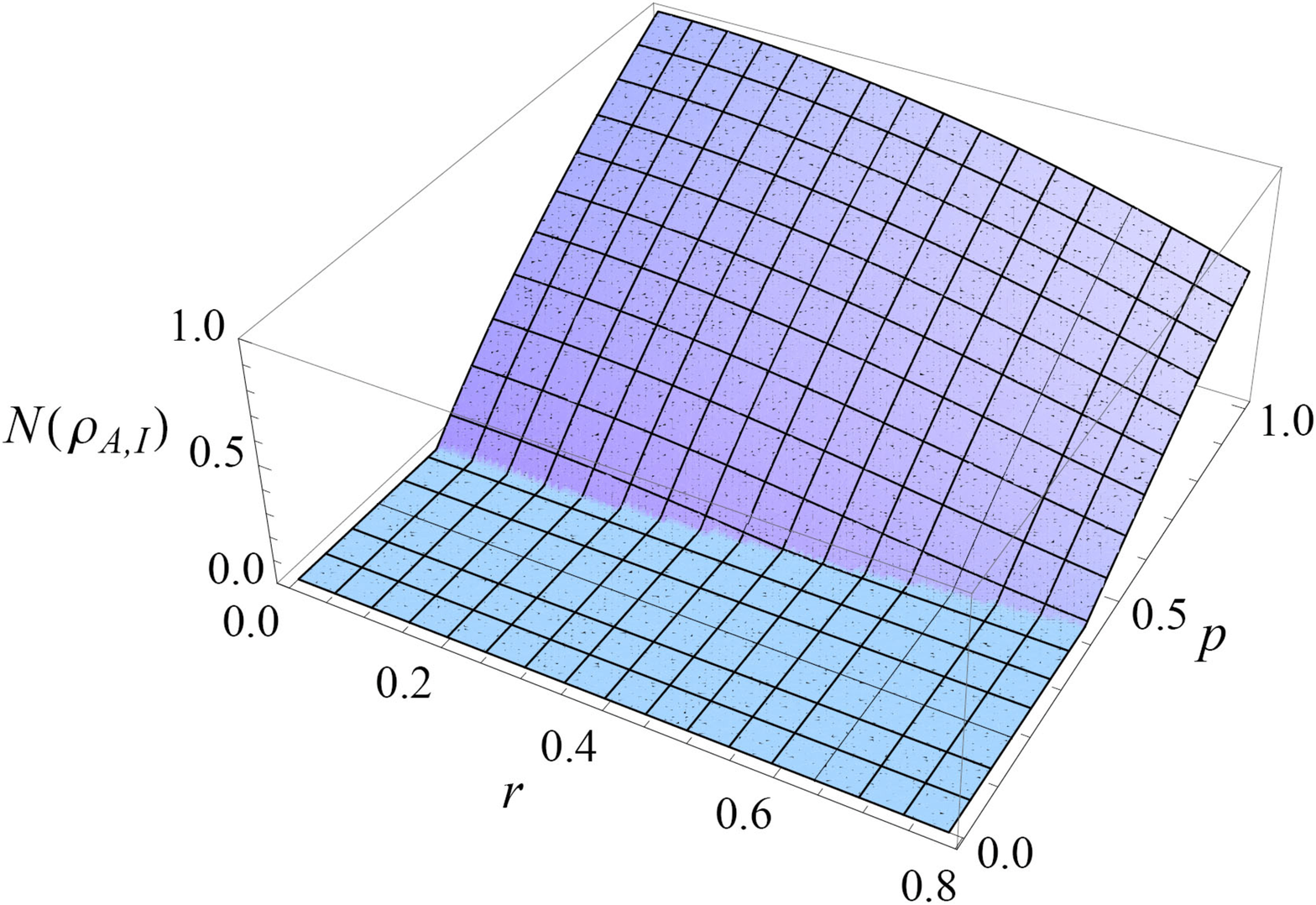}
    \parbox{15cm}{\caption{  Logarithmic negativity,
$N(\rho_{A,\tI})$ as funnction of $p$ and $r$.
    \label{fig2}}}\end{center}
    \end{figure}

For quantifying entanglement,  the logarithmic negativity is
evaluated for $\rho_{A,\tI}$. Firstly, the eigenvalues of
$\rho_{A,\tI}^{\rm pt}$ are evaluated as follows
    \bea
    &&\lambda_{1,2}(\rho_{A,\tI}^{\rm pt})=\frac{1}{4}
    \left\{1-p\cos^2r\pm\sqrt{\sin^4r+4p^2\cos^2r}\right\},\nn\\
    &&\lambda_{3,4}(\rho_{A,\tI}^{\rm
    pt})=\frac{1}{4}\left\{1+p\cos^2r\pm\sin^2r\right\}.
    \label{lambda-t-AI}
    \eea
Using Eq. (\ref{logneg}),  the logarithmic negativity,
$N(\rho_{A,\tI})$, is evaluated and plotted as a function of
acceleration, $r$, and fraction rate, $p$, in Fig. \ref{fig2}. It is
clear that the entanglement of the state degrades with increasing
the acceleration.

It should be mentioned that all the eigenvalues, Eq.
(\ref{lambda-t-AI}), are positive except for the one appearing on
the first line with a minus sign, {\it i.e.},
$\lambda_2(\rho_{A,\tI}^{\rm
pt})=\left(1-p\cos^2r-\sqrt{\sin^4r+4p^2\cos^2r}\right)/4$. In the
inertial limit, $r=0$, this eigenvalue is negative for $p>1/3$, and
it is an increasing function of $r\in[0,\pi/4]$. In the limit of
infinite acceleration, $r=\pi/4$, $\lambda_2(\rho_{A,\tI}^{\rm pt})$
is negative for $p>3/7$. In summary, for the region $p\in(1/3,3/7)$,
whose inertial state is entangled, one may find some finite
acceleration that makes the noninertial state separable.

As far as here the entanglement is studied for bipartite states the logarithmic negativity is enough to be evaluated. However, one may imply other measures of entanglement for investigation of the above states. One of the existing entanglement measure to be evaluated is entanglement of formation. The entanglement of formation is
defined as \cite{wootters}
 \bea
 E_f(\rho)=-\frac{1+\sqrt{1-C^2}}{2}\log_2
 \frac{1+\sqrt{1-C^2}}{2}-\frac{1-\sqrt{1-C^2}}{2}\log_2
 \frac{1-\sqrt{1-C^2}}{2},
 \eea
 where
$$C(\rho):=\max\left\{0,\tilde{\lambda}_1-\tilde{\lambda}_2-\tilde{\lambda}_3
-\tilde{\lambda}_4\right\},~~\tilde{\lambda}_i\geq\tilde{\lambda}_{i+1}\geq
0.$$ Here $\tilde{\lambda}_i$ are defined by the square roots of
 (positive) ordered eigenvalues of the operator
$\rho(\sigma_y\otimes \sigma_y)\rho^*(\sigma_y\otimes \sigma_y)$.
  \begin{figure}[t]
    \begin{center}
   \includegraphics[scale=.17,clip]{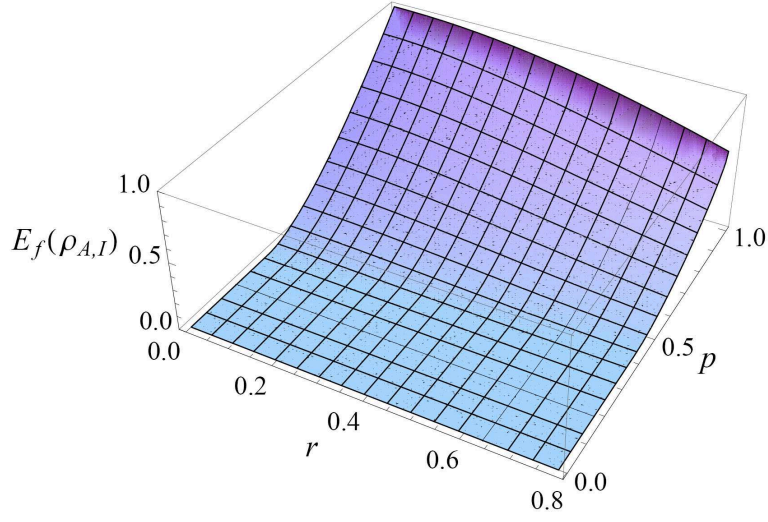}
    \parbox{15cm}{\caption{ Entanglement of formation, $E_f(\rho_{A,\tI})$ as
  function of $p$ and $r$.
    \label{fig8}}}\end{center}
    \end{figure}
Figure \ref{fig8} represents the entanglement of formation for
$\rho_{A,\tI}$ and shows that the state is entangled for $p>3/7$ at
infinite acceleration limit. The general behavior of this measure
and the one for logarithmic negativity are the same, confirming our expectation, therefore, in the
following we continue evaluating only the logarithmic negativity.

\subsection{Bipartition Alice-AntiRob}
Tracing  out over the Rob's modes of $\rho_{A,\tI,\tII}$,  the
relating density matrix for the bipartite Alice and antiRob is
calculated as
     \bea
      \rho_{A,\tII}=\frac{1}{4}\left(
                   \begin{array}{cccc}
                     1+\cos^2 r-p\sin^2 r & 0 & 0 & 0 \\
                     0 & (1+p)\sin^2 r &2p \sin r & 0 \\
                     0 &2p \sin r & 1+\cos^2r+p\sin^2r & 0 \\
                      0& 0 & 0 &(1-p)\sin^2 r \\
                   \end{array}
                 \right),
    \label{rho-AII}
    \eea
where, we have used the basis $|0\kt_A|0\kt_{\tII}$,
$|0\kt_A|1\kt_{\tII}$, $|1\kt_A|0\kt_{\tII}$, and
$|1\kt_A|1\kt_{\tII}$. The eigenvalues of $\rho_{A,\tII}$  are given
by
    \bea
    \lambda_{1,2}(\rho_{A,\tII})&=&\frac{1}{4}\left\{1+p\sin^2r\pm\sqrt{\cos^4r+4p^2\sin^2r}\right\},\nn\\
    \lambda_{3,4}(\rho_{A,\tII})&=&\frac{1}{4}\left\{1-p\sin^2r\pm\cos^2r\right\}.
    \eea
Using these eigenvalues,  the von Neumann entropy of $\rho_{A,\tII}$
is given by $S(\rho_{A,\tI})=-\sum_{j=1}^4
\lambda_j\log_2\lambda_j$. Similarly  the entropies of the reduced
density matrices $\rho_A$ and $\rho_{\tI}$ are obtained  as
$S(\rho_A)=1$ and
$S(\rho_{\tII})=-\frac{\sin^2r}{2}\log_2\frac{\sin^2r}{2}-\frac{1+\cos^2r}{2}\log_2\frac{1+\cos^2r}{2}$,
respectively.

 \begin{figure}[t]
    \begin{center}
   \includegraphics[scale=.17,clip]{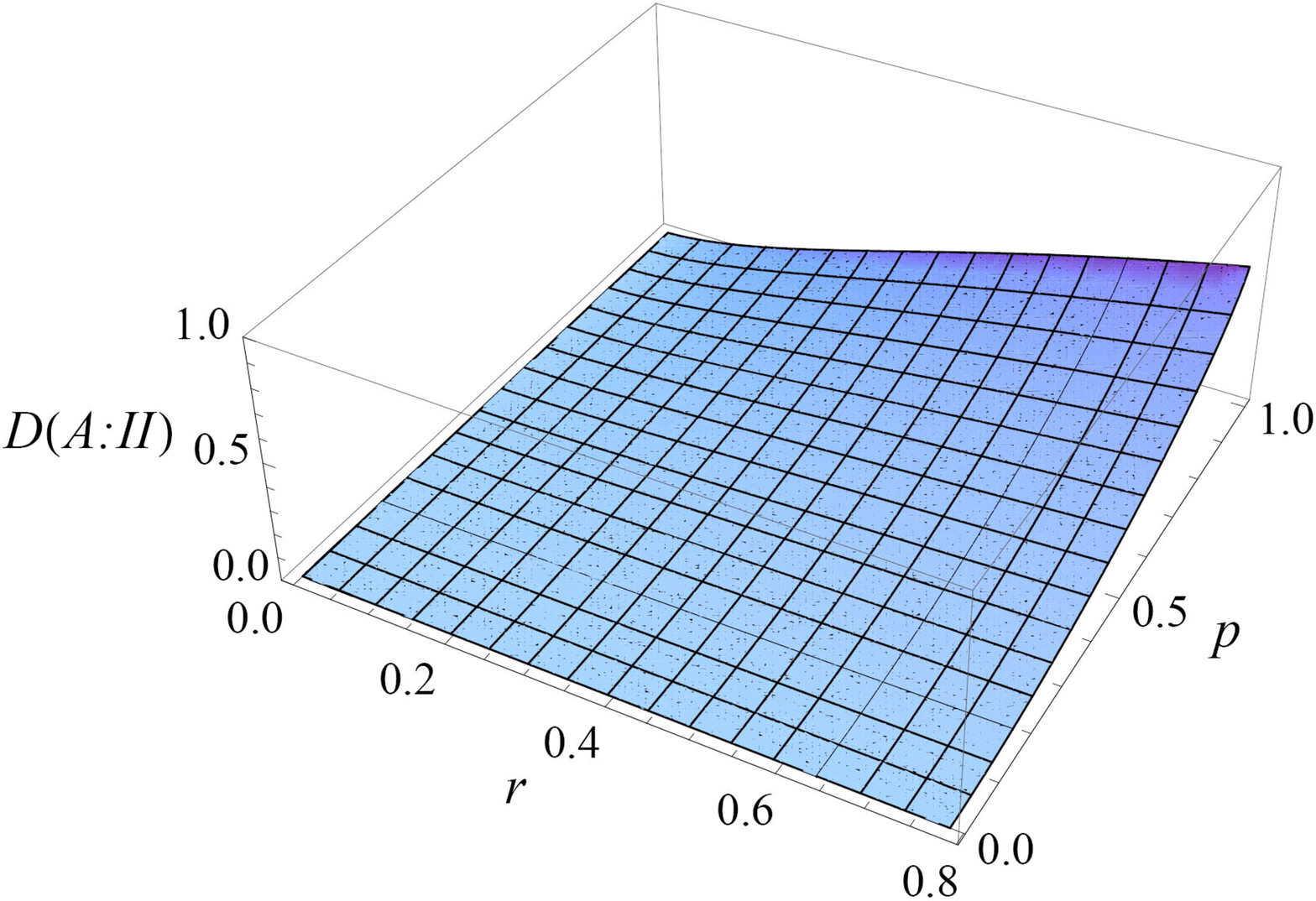}
    \parbox{15cm}{\caption{ Quantum discord, $\cD(A:\tII)$, as function of $r$
and $p$.
    \label{fig3}}}\end{center}
    \end{figure}

Thus, quantum discord is found to be as follows
    \bea
    \cD(A:\tII)=\min_{\{\Pi_k\}}[S_{\{\Pi_k\}}(A|\tII)]-S(\rho_{A,\tII})
    -\mbox{$\frac{1+\cos^2r}{2}$}\log_2\mbox{$\frac{1+\cos^2r}{2}$}-
    \mbox{$\frac{\sin^2r}{2}$}\log_2\mbox{$\frac{\sin^2r}{2}$}.
    \label{disAII}
    \eea
It can be shown that $S_{\{\Pi_k\}}(A|\tII)$ attains its minimum in
the case $\kappa=\ell=1/2$ with $\beta=1/4$ (for more details see
the Appendix). Using the evaluated
$\min_{\{\Pi_k\}}S_{\{\Pi_k\}}(A|\tII)$, quantum discord is plotted
in Fig. \ref{fig3}. It is clear that quantum discord is an
increasing function of acceleration. Indeed, $\cD(A,\tII)$ is zero
for $r=0$ and it is an increasing function of acceleration (except
for the case $p=0$, at which it is permanently equal to zero) and
tends toward the value of $\cD(A,\tI)$ in the limit of infinite
acceleration.
 \begin{figure}[t]
    \begin{center}
   \includegraphics[scale=.17,clip]{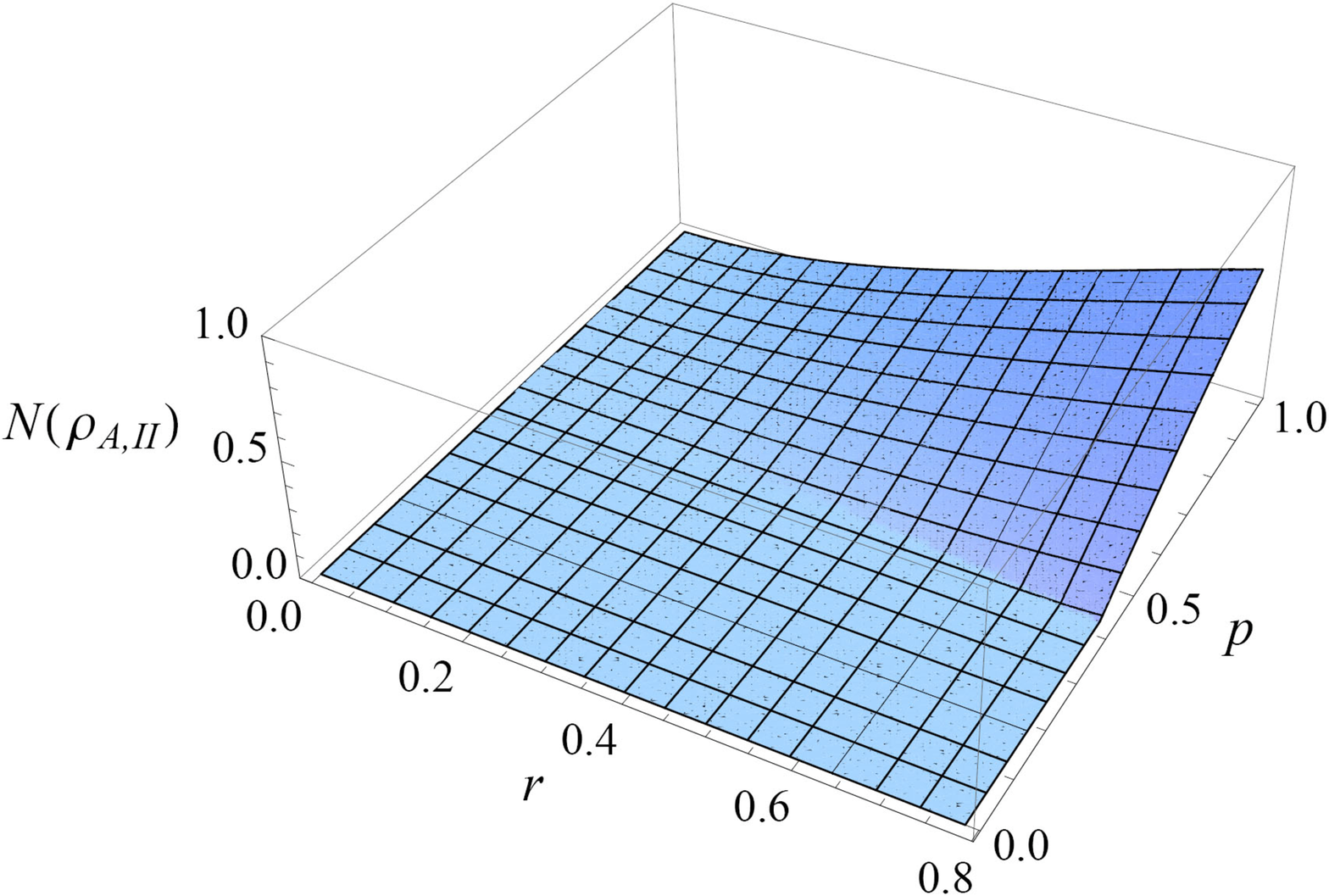}
    \parbox{15cm}{\caption{ Logarithmic negativity, $N(\rho_{A,\tII})$ as
function of $p$ and $r$.
    \label{fig4}}}\end{center}
    \end{figure}

The eigenvalues of $\rho_{A,\tII}^{\rm pt}$ are given by
    \bea
    \lambda_{1,2}(\rho_{A,\tII}^{\rm pt})&=&\frac{1}{4}\left\{1-p\sin^2r\pm\sqrt{\cos^4r+4p^2\sin^2r}\right\},\nn\\
    \lambda_{3,4}(\rho_{A,\tII}^{\rm pt})&=&\frac{1}{4}\left\{1+p\sin^2r\pm\cos^2r\right\}.
    \label{lambda-t-AII}
    \eea
Thus, the logarithmic negativity is calculated and plotted in Fig.
\ref{fig4}. It is clear for $r=0$, the pseudo-entangled state is
separable for all values of $p$. By increasing the acceleration, the
state for large values of $p$ becomes entangled. for $r=\pi/4$, the
sate is entangled for $p\in[3/7,1]$, while it remains nonentangled
for $p<3/7$. Similar to the previous case, one can find that  the
second eigenvalue, $\lambda_2(\rho_{A,\tII}^{\rm
pt})=\left(1+p\sin^2r-\sqrt{\cos^4r+4p^2\sin^2r}\right)/4$, is
negative for  finite values of acceleration for $p$ greater than a
specific value. In the infinite acceleration limit, this eigenvalue
is negative for $p>3/7$. Unlike the previous case,  the entanglement
of the state upgrades with respect to acceleration. Also, it can be
verified that in the infinite limit of acceleration, the logarithmic
negativity of the pseudo-entangled state in Alice-Rob and
Alice-antiRob bipartitions completely coincide with each other.

\subsection{Bipartition Rob-AntiRob}
 By tracing out Alice, the density matrix of Rob and antiRob is given
 by
    \bea
    \rho_{\tI,\tII}=\frac{1}{2}\left(
                   \begin{array}{cccc}
                     \cos^2 r & 0 & 0 & \sin r \cos r \\
                     0 & 0 & 0 & 0 \\
                     0 & 0 & 1 & 0 \\
                     \sin r \cos r & 0 & 0 & \sin^2r \\
                   \end{array}
                 \right),
    \label{rho-III}
    \eea
where, we have used the basis $|0\kt_{\tI}|0\kt_{\tII}$,
$|0\kt_{\tI}|1\kt_{\tII}$, $|1\kt_{\tI}|0\kt_{\tII}$, and
$|1\kt_{\tI}|1\kt_{\tII}$. This is exactly same as the density
matrix obtained for the special case $p=1$ \cite{ref06,ref11}. Since
the entanglement of $\rho_{\tI,\tII}$ has been studied elsewhere
\cite{ref06,ref11,ref53}, it will not be presented here. The quantum
discord, $\cD(\tI:\tII)$, is evaluated and plotted in Fig.
\ref{fig5}. It is shown that $\cD(\tI:\tII)$ is an increasing
function of acceleration.
\subsection{The Local Unitary Equivalent States}
One can construct the equivalent state for each states, such as the
pseudo-entangled state studied in this paper, by applying local unitary operations. By operating a local
operation such as $\sigma_{x}$ on the second partition of Eq.
(\ref{pseu-ent-gen}) we have
 \bea
 \tilde{\rho}=\frac{1-p}{4}I+p|\Psi^+\kt\br\Psi^+|,
 \eea
where, $|\Psi^+\kt=\frac{1}{\sqrt{2}}(|01\kt+|10\kt)$. It is an
elementary exercise to check that the logarithmic negativity of
$\rho$ and $\tilde{\rho}$ is the same and therefore, these two
states are called the equivalent states in an inertial frame. In
previous section, we considered the entanglement of $\rho$ in a
noninertial frame, hence, we can do the same for
$\tilde{\rho}$. Although these two density matrices are different in
noninertial frame but their partial transposes have the same
eigenvalues and therefore, the same logarithmic negativity. In other
words, these two equivalent states still remain equivalent in a
noninertial frame. It is mentionable that this is not a general
property.

\begin{figure}[t]
    \begin{center}
   \includegraphics[scale=.2,clip]{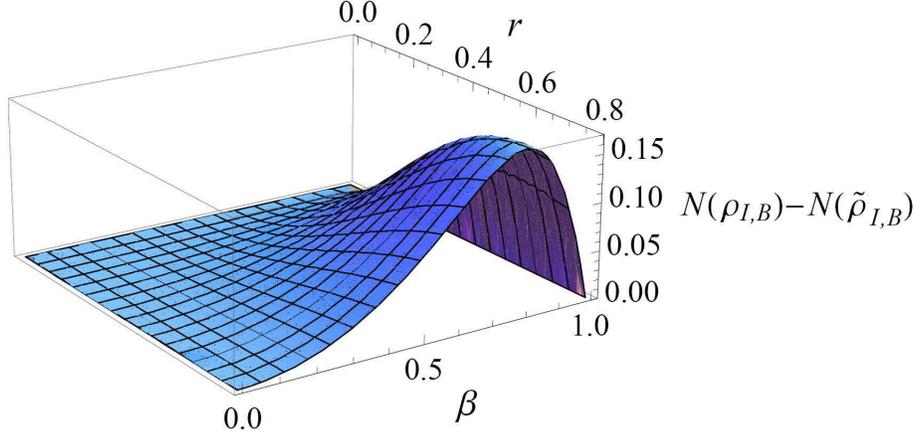}
    \parbox{15cm}{\caption{ Logarithmic negativity, $N(\rho_{\tI,B})-N(\tilde{\rho}_{\tI,B})$ as
function of $p$ and $r$.
    \label{fig7}}}\end{center}
    \end{figure}

We can investigate the behavior of
$|\psi\kt=\alpha(|00\kt+|11\kt)+\beta|10\kt$ and it's equivalent
$|\tilde{\psi}\kt=(\sigma_x \otimes {I})
|\psi\kt=\alpha(|10\kt+|01\kt)+\beta|00\kt$ with
$2\alpha^{2}+\beta^{2}=1$, in a noninertial frame. If the first parties
of these states are measured in an accelerating frame, similar to the
previous section we can obtain $\rho_{\tI,\tII,B}$
(${\tilde{\rho}_{\tI,\tII,B}}$) and by tracing over the region $\tII$, we have the following states.
 \bea
    \rho_{\tI,B}=\left(
                   \begin{array}{cccc}
                     \alpha^2 \cos^2 r & 0 & \alpha \beta \cos r & \alpha \cos^2 r  \\
                     0 & 0 & 0 & 0 \\
                    \alpha \beta \cos r  & 0 & \beta^2 + \alpha^2 \sin^2r & \alpha \beta \\
                      \alpha \cos^2 r  & 0 & \alpha \beta & \alpha^2 \\
                   \end{array}
                 \right),
    \label{rho-IB}
    \eea
 \bea
   \tilde{ \rho}_{\tI,B}=\left(
                   \begin{array}{cccc}
                     \beta^2 \cos^2 r &\alpha\beta \cos^2 r  & \alpha \beta \cos r & 0  \\
                     \alpha\beta \cos^2 r & \alpha^2 \cos^2 r & \alpha^2 \cos r & 0 \\
                    \alpha \beta \cos r  & \alpha^2 \cos r &  \alpha^2+\beta^2 \sin^2r & \alpha \beta\sin^2r \\
                      0  & 0 & \alpha \beta\sin^2r & \alpha^2\sin^2r \\
                   \end{array}
                 \right).
    \label{rho-t-IB}
    \eea
Figure \ref{fig7} shows the differences between the logarithmic
negativity of these initially equivalent states in a noninertial
frame. The special cases ($\beta=0$ and $\beta=1$) which represent
the maximally entangled and separable states, respectively, still are equivalent in
noninertial frame too. However, one can find a special case with
$\beta\approx0.80$ which, face with maximum differences between
initial equivalent states in the infinite acceleration limit.

Finding a general description on the behavior of equivalent classes of states in noninertial frames requires further investigation and it will be explored somewhere else.

\section{Discussion and Conclusion }
We studied the status of entanglement and nonclassical correlations
for the pseudo-entangled state (\ref{pseu-ent-gen}), by evaluating
logarithmic negativity, $N$, and quantum discord, $\cD$,
respectively.  $p$ is a representation of polarization in the
physical systems such as NMR and ENDOR. In an inertial frame, if $p
\leq p_{\rm c}$, where $p_{\rm c}=1/3$, there is no entanglement in
$\rho$; however nonclassical correlations may still exist among the
subsystems of $\rho$.

For the case when correlations between Alice and Rob,
$\rho_{A,\tI}$, are studied, it is shown that $N(\rho_{A,\tI})$ is
generally decreasing as a function of acceleration and it reaches
its minimum $N_{\rm min}(\rho_{A,\tI})$ for an infinite
acceleration. It is also found that $p_{c}$ is increasing from
$p_{\rm c}=1/3$ for an inertial frame up to $p_{\rm c}=3/7$ for an
infinite acceleration, $r=\pi/4$. This means that there is a
critical range in which entanglement is destroyed due to
acceleration. In other words, in noninertial frames, it can be more
challenging to produce entanglement as compared to an inertial frame
if $p$ falls into the critical range. Quantum discord, $\cD(A:\tI)$,
decreases from a nonzero value, for $p>0$, with respect to
acceleration and it reaches its minimum $\cD_{\rm min}(A:\tI)$ for
$r=\pi/4$. There is no critical range detected for discord in which
its behavior is similar to that for entanglement.

Similar arguments are put forward for the partition Alice and
antiRob, $\rho_{A,\tII}$. For the stationary case, $r=0$,
$N(\rho_{A,\tII})=\cD(A:\tII)=0$, regardless of $p$. With slightly
increasing acceleration, and for $3/7<p \leq 1$, $N(\rho_{A,\tII})$
increases up to its maximum $N_{\rm max}(\rho_{A, \tII})$ for
$r=\pi/4$. The behavior of  quantum discord is simple to explain in
this case as it starts increasing as a function of $r$ and for any
$p> 0$. It reaches  its maximum value, $\cD_{\rm max}(A:\tII)$, in
the limit of infinite acceleration .

It is noteworthy that for $r=\pi/4$, the logarithmic negativities of
$\rho_{A,\tI}$ and $\rho_{A,\tII}$ approach to each other, i.e,
$N_{\rm min}(\rho_{A,\tI})=N_{\rm max}(\rho_{A,\tII})$. This result
is consistent with what has been reported for the special case of
maximally entangled state \cite{ref06}. Similarly, $\cD_{\rm
min}(A:\tI)=\cD_{\rm max}(A:\tII)$. This has been reported, in the
special case of maximally entangled state, by \cite{ref11}.

From the studies on the maximally entangled state, a conjecture
might be that entanglement is conserved among the bipartions,
Alice-Rob and Alice-antiRob. Meaning that, e.g., once the
entanglement of Alice-Rob is decreasing, followingly the
entanglement of Alice-antiRob is increasing. Our general study shows
that entanglement cannot be conserved since at least there is a
region in which the entanglement of Alice-Rob vanishes whereas the
entanglement of Alice-antiRob never increases but stays in its
original zero value. Note that it is clear from Eq. (\ref{rho-III})
that the entanglement of Rob and antiRob is independent of $p$.

 \begin{figure}[t]
    \begin{center}
   \includegraphics[scale=.75,clip]{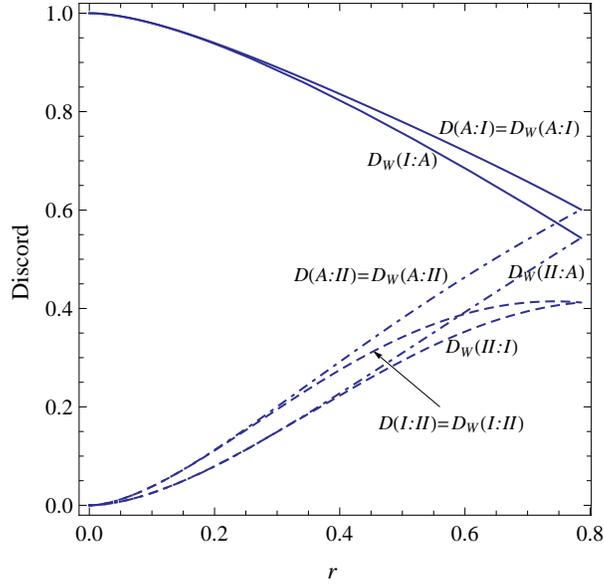}
    \parbox{15cm}{\caption{ Quantum discords
    for bipartitions from Alice, Rob and antiRob as functions of $r$.
    \label{fig5}}}\end{center}
    \end{figure}

Wang {\it et. al.} \cite{ref11} picked up a maximally entangled
state for their study. The state can be considered as a special case
of the class of states studied in this work. Therefore, it is
expected that the results of our work, for $p=1$, match with those
reported in \cite{ref11}. Following the method employed in
\cite{ref11}, we have calculated $\cD_{\rm W}(\tI:A)$, $\cD_{\rm
W}(\tII:A)$, and $\cD_{\rm W}(\tII:\tI)$. The corresponding values
are shown in Fig. \ref{fig5}. Employing the approach developed in
this work, $\cD(A:\tI)$, $\cD(A:\tII)$, and $\cD(\tI:\tII)$ are
calculated and shown in Fig. \ref{fig5}. Clearly, the corresponding
values of $\cD$ for each bipartition do not match to each other.
This inconsistency is however resolved by a simple explanation. In
\cite{ref11} the von Neumann projectors are applied to the first
parts of bipartitions, whereas we have applied the projectors on the
second parts. If the von Neumann projectors of \cite{ref11} are
applied to the second parts of bipartitions then
$\cD(A:\tI)=\cD_{\rm W}(A:\tI)$, $\cD(A:\tII)=\cD_{\rm W}(A:\tII)$,
and $\cD(\tI:\tII)=\cD_{\rm W}(\tI:\tII)$. Hence, a perfect
consistency is observed between the results of \cite{ref11}, and
those of ours, see Fig. \ref{fig5}.

Fig. \ref{fig5}, is also a nice example to show the nonsymmetric
behavior of quantum discord  mentioned in section II as originally
discussed by Zurek \cite{ref47}.

\section*{Acknowledgment}
We would like to acknowledge one the referees for his/her valuable
comments about the behavior of equivalent states in noninertial
frames. We would like to acknowledge Tahmineh Godazgar for her
assistance in preparing the first draft of this paper, and Ivette
Fuentes for useful discussions. H. M.-D. would like to acknowledge
Takeji Takui for hospitality, during his visits in Osaka City
University. H. M.-D. is supported by the ``Open Research Center''
Project for Private Universities: matching fund subsidy from MEXT.
R. R. is supported by JSPS through FIRST Program, also Industry of Canada and CIFAR.
 \section*{Appendix: Discord for Real Symmetric $X$-State }
Ali {\it et. al.}  have evaluated \cite{ref52} the quantum discord
of general Hermitian $X$-shaped states represented by
    \bea
    \rho_{A,B}=\left(
           \begin{array}{cccc}
             \rho_{11} & 0 & 0 & \rho_{14} \\
             0 & \rho_{22} & \rho_{23} & 0 \\
             0 & \rho_{32} & \rho_{33} & 0 \\
             \rho_{41} & 0 & 0 & \rho_{44} \\
           \end{array}
         \right).
    \eea
For convenience, a brief review of Ref. \cite{ref52} for a special
real symmetric case, $\rho=\rho^*=\rho^\dag$, is given here, which
will be used for the calculations presented in this paper.

The main obstacle in evaluating quantum discord, $\cD(A:B)$,  is to
obtain the minimum value of $S_{\{\Pi_k\}}(A|B)$ over all possible
von Neumann measurements, {\it i.e.},
    \bea
    \Pi_k=I \otimes V |k\kt\br k|V^{\dag},~~~~~~k=0,1.
    \label{proj-gen}
    \eea
Here, $V$ is a general $SU(2)$ element that can be written as
\cite{ref36,ref52}
    \bea
    V=tI+i\vec{y}.\vec{\bf{\sigma}}, ~~~~~~t,y_1,y_2,y_3\in \RR,
    \label{V}
    \eea
with $t^2+\vec{y}.\vec{y}=1$. By performing measurement $\Pi_k$, the
state collapses to the density matrix $\rho_{A|k}$ with the
probability $p_k$. One can easily find $p_k$ as
    \bea
    p_0=1-p_1=(\rho_{22}+\rho_{44})\ell+(\rho_{11}+\rho_{33})\kappa,
    \label{p-i}
    \eea
with $\ell=1-\kappa=y_1^2+y_2^2\in[0,1]$. The eigenvalues of
$\rho_{A|k}$ can be obtained \cite{ref52} as
    \bea
    \lambda_{\pm}(\rho_{A|k})=\frac{1}{2}(1\pm \theta_k),
    \label{lambda-t}
    \eea
where, $\theta_k$ are given by
    \bea
    \theta_0&=&\frac{1}{p_0}\sqrt{[(\rho_{11}-\rho_{33})\kappa
    +(\rho_{22}-\rho_{44})\ell]^2+\beta},\label{theta-k0}\\
    \theta_1&=&\frac{1}{p_1}\sqrt{[(\rho_{11}-\rho_{33})\ell
    +(\rho_{22}-\rho_{44})\kappa]^2+\beta} ,
    \label{theta-k}
    \eea
where, $\beta=4\kappa \ell (\rho_{14}+\rho_{23})^2-16
\mu\rho_{14}\rho_{23}$, with $\mu=(ty_1+y_2y_3)^2\in[0,1/4]$. Having
determined the eigenvalues of $\rho_{A|k}$, one can easily evaluate
$S_{\{\Pi_k\}}(A|B)$ along the following line
    \bea
    S_{\{\Pi_k\}}(A|B)=p_0S(\rho_{A|0})+p_1S(\rho_{A|1}),
    \label{SAB-xshaped}
    \eea
where, $S(\rho_{A|k})=-\sum_{\pm}\lambda_{\pm}(\rho_{A|k})\log_2
\lambda_{\pm}(\rho_{A|k})$.

It is clear that $S_{\{\Pi_k\}}(A|B)$ is an even function of
$\kappa-\ell$ and, hence, it attains a minimum value at the middle
point $\kappa=\ell=1/2$ or at the end points $\kappa=1-\ell=0,1$ (at
which $\mu=0$) \cite{ref52}. In the case of $\kappa=1/2$,
$S(\rho_{A|0})=S(\rho_{A|1})$ and the minimization of
$S_{\{\Pi_k\}}(A|B)$ is equivalent to the minimization of
$S(\rho_{A|0})$. Since $\beta$ is a linear function of $\mu$, it can
be shown that $\theta_0$, and thereby $S(\rho_{A|0})$ attain their
minimum values at one of the end points $\mu=0,1/4$. Now, by
checking all the above mentioned cases,
$\min_{\{\Pi_k\}}[S_{\{\Pi_k\}}(A|B)]$ and, consequently, quantum
discord can be calculated using Eqs. (\ref{J}-\ref{discord}).


\begin{thebibliography}{99}

\bibitem{ref01}
A. Peres , and D. R. Terno, Rev. Mod. Phys. {\bf 76}, 93 (2004).

\bibitem{ref02}
E. Schr\"{o}dinger, Naturwissenschaften, {\bf  23}, 807 (1935);
ibid. {\bf 23}, 823 (1935); ibid. {\bf 23}, 844 (1935).

\bibitem{ref03}
R. Horodecki, P. Horodecki, M. Horodecki, and K. Horodecki, Rev.
Mod. Phys. {\bf 81}, 865 (2009).

\bibitem{ref04}
P. M. Alsing, and G. J. Milburn, Phys. Rev. Lett. {\bf 91}, 180404
(2003).

\bibitem{ref05}
I. Fuentes-Schuller, and R. B. Mann, Phys, . Rev. Lett. {\bf 95},
120404 (2005).

\bibitem{ref06}
P. M. Alsing, I. Fuentes-Schuller, R. B. Mann, and T. E. Tessier,
Phys. Rev. A {\bf 74}, 032326 (2006).

\bibitem{ref07}
J. Le\'{o}n, and E. Mart\'{i}n-Mart\'{i}nez, Phys. Rev. A {\bf 80},
012314 (2009).

\bibitem{ref08}
R. B. Mann, and V. M. Villalba, Phys. Rev. A {\bf 80}, 022305
(2009).

\bibitem{ref09}
S. D. Bartlett, T. Rudolph, and R. W. Spekkens, Rev. Mod. phys. {\bf
79}, 555 (2007).

\bibitem{ref10}
E. Mart\'{i}n-Mart\'{i}nez, and J. Le\'{o}n, Phys. Rev. A {\bf 81},
032320 (2010).

\bibitem{ref11}
J. Wang, J. Deng, and J. Jing, Phys. Rev. A {\bf 81}, 052120 (2010).

\bibitem{ref12}
M. Aspelmeyer, H. R. B\"{o}hm, T. Gyatso, T. Jennewein, R.
Kaltenbaek, M. Lindenthal, G. Molina-Terriza, A. Poppe, K. Resch, M.
Taraba, R. Ursin, P. Walther, and A. Zeilinger, Science {\bf 301},
621 (2003).

\bibitem{ref13}
C.-Z. Peng, T. Yang, X.-H. Bao, J. Zhang, X.-M. Jin, F.-Y. Feng, B.
Yang, J. Yang, J. Yin, Q. Zhang, N. Li, B.-L. Tian, and J.-W. Pan,
Phys. Rev. Lett. {\bf 94}, 150501 (2005).

\bibitem{ref14}
R. Ursin, F. Tiefenbacher, T. Schmitt-Manderbach, H. Weier, T.
Scheidl, M. Lindenthal, B. Blauensteiner, T. Jennewein, J.
Perdigues, P. Trojek, B. \"{O}mer, M. F\"{u}rst, M. Meyenburg, J.
Rarity, Z. Sodnik, C. Barbieri, H. Weinfurter, and  A. Zeilinger,
Nature Phys. {\bf 3}, 481 (2007).

\bibitem{ref15}
R. Ursin, T. Jennewein, J. Kofler, J. M. Perdigues, L. Cacciapuoti,
C. J. de Matos, M. Aspelmeyer, A. Valencia, T. Scheidl, A. Fedrizzi,
A. Acin, C. Barbieri, G. Bianco, C. Brukner, J. Capmany, S. Cova, D.
Giggenbach, W. Leeb, R. H. Hadfield, R. Laflamme, N. Lutkenhaus, G.
Milburn, M. Peev, T. Ralph, J. Rarity, R. Renner, E. Samain, N.
Solomos, W. Tittel, J. P. Torres, M. Toyoshima, A. Ortigosa-Blanch,
V. Pruneri, P. Villoresi, I. Walmsley, G Weihs, H. Weinfurter, M.
Zukowski, and A. Zeilinger, Proc. Int. Astronaut. Congr. A2.1.3
(2008), arXiv:0806.0945.

\bibitem{ref16}
A. Fedrizzi, R. Ursin, T. Herbst, M. Nespoli, R. Prevedel, T.
Scheidl, F. Tiefenbacher, T. Jennewein, and A. Zeilinger, Nature
Phys. {\bf 5}, 389 (2009).

\bibitem{ref17}
P. C. W. Davies, J. Phys. A: Math. Gen. {\bf 8}, 609 (1975); W. G.
Uruh. Phys. Rev. D {\bf 14}, 870 (1976).

\bibitem{ref18}
H. Ollivier, and W. H. Zurek, Phys. Rev. Lett. {\bf 88}, 017901
(2001).

\bibitem{ref19}
L. Henderson, and V. Vedral, J. Phys. A: Math. Gen. {\bf 34}, 6899
(2001).

\bibitem{ref20}
V. Vedral, Phys. Rev. Lett. {\bf 90}, 050401 (2003).

\bibitem{ref21}
A. Datta, A. Shaji, and C. M. Caves, Phys. Rev. Lett. {\bf 100},
050502 (2008).

\bibitem{ref22}
J. A. Jones, M. Mosca, and R. H. Hansen, Nature {\bf 393}, 344
(1998).

\bibitem{ref23}
R. Rahimi, K. Sato, K. Furukawa, K. Toyota, D. Shiomi, T. Nakamura,
M. Kitagawa, and T. Takui, Int. J. Quantum Inf. {\bf 3}, 197,
(2005).

\bibitem{ref24}
 D. P. Divincenzo, Forschr. Phys.  {\bf 48}, 771 (2000).

\bibitem{ref25}
R. F. Werner, Phys. Rev. A {\bf 40}, 4277 (1989).

\bibitem{ref26}
 A. Peres, Phys. Rev. Lett. {\bf 77}, 1413 (1996).

\bibitem{ref27}
 M. B. Plenio, and S. Virmani, Quant. Inf. Comp. {\bf 7}, 1 (2007).

\bibitem{ref28}
C. H. Bennett, D. P. DiVincenzo, C. A. Fuchs, T. Mor, E. Rains, P.
W. Shor, J. A. Smolin, and W. K. Wootters, Phys. Rev. A {\bf 59},
1070 (1999).

\bibitem{ref29}
J. Oppenheim, M. Horodecki, P. Horodecki, and R. Horodecki, Phys.
Rev. Lett. {\bf 89}, 180402 (2002).

\bibitem{ref30}
B. Groisman, S. Popescu, and A. Winter, Phys. Rev. A {\bf 72},
032317 (2005).

\bibitem{ref31}
M. Horodecki, P. Horodecki, R. Horodecki, J. Oppen-heim, A. Sen(De),
U. Sen, and B. Synak-Radtke , Phys. Rev. A {\bf 71}, 062307 (2005).

\bibitem{ref32}
M. B. Plenio, and S. Virmani, Quant. Inf. Comp. {\bf 7}, 1 (2007).

\bibitem{ref33}
M. B. Plenio, Phys. Rev. Lett. {\bf 95}, 090503 (2005).

\bibitem{ref34}
B. Groisman, D. Kenigsberg, and T. Mor, arXiv:quant-ph/0703103.

\bibitem{ref35}
A. SaiToh, R. Rahimi, and M. Nakahara, Phys. Rev. A {\bf 77}, 052101
(2008).

\bibitem{ref36}
S. Luo, Phys. Rev. A {\bf 77}, 042303 (2008).

\bibitem{ref37}
A. Datta, and S. Gharibian, Phys. Rev. A {\bf 79}, 042325 (2009).

\bibitem{ref38}
 A. SaiToh, R. Rahimi, and M. Nakahara, arXiv: 0802.2263.

\bibitem{ref39}
A. SaiToh, R. Rahimi, and M. Nakahara, arXiv: 0906.4187.

\bibitem{ref40}
 R. Rahimi, and A. SaiToh, Phys. Rev. A {\bf 82}, 022314 (2010).

\bibitem{ref41}
K. Modi, T. Paterek, W. Son, V. Vedral, and M. Williamson, Phys.
Rev. Lett. {\bf 104}, 080501 (2010).

\bibitem{ref42}
J. Cui, and H. Fan, J. Phys. A: Math. Theor. {\bf 43}, 045305
(2010).

\bibitem{ref43}
M. Piani, P. Horodecki, and R. Horodecki, Phys. Rel. Lett. {\bf
100}, 090502 (2008).

\bibitem{ref44}
 M. Piani, M. Christandl, C. E. Mora, and P. Horodecki, Phys. Rev. Lett, {\bf 102}, 250503
 (2009).

\bibitem{ref45}
 A. Ferraro, L. Aolita, D. Cavalcanti, F. M. Cucchietti, and A. Acin, arXiv:
 0908.3157.

\bibitem{ref46}
J. Maziero, L. C. C\'{e}leri, R. M. Serra, and V. Vedral,  Phys.
Rev. A {\bf 80}, 044102 (2009).

\bibitem{ref47}
W. H. Zurek, Phys. Rev. A {\bf 67}, 012320 (2003).

\bibitem{ref48}
S. Takagi, Prog. Theor. Phys. Suppl. {\bf 88} 1 (1986).

\bibitem{ref49}
R. J\'{a}uregui, M. Torres, and S. Hacyan, Phys. Rev. D {\bf 43},
3979 (1991).

\bibitem{ref50}
M. Soffel, B. M\"{u}ller, and W. Greiner, Phys. Rev. D {\bf 22},
1935 (1980).
\bibitem{ref51}
D. E. Bruschi, J. Louko, E. Mart\'{i}n-Mart\'{i}nez, A. Dragan, and
I. Fuentes, arXiv:1007.4670.

\bibitem{wootters} W. K. Wootters, Phys.\ Rev.\ Lett. {\bf 80}, 2245 (1998).


\bibitem{ref53}
 S.~Moradi, Phys.\ Rev.\ A {\bf 79}, 064301 (2009).

\bibitem{ref52}
 M. Ali, A. R. P. Rau, and G. Alber, Phys. Rev. A {\bf 81}, 042105
(2010).

\end{thebibliography}
\end{document}